\documentclass[usenatbib]{aa}

\usepackage{txfonts}
\usepackage{hyperref}
\usepackage{graphicx}   
\usepackage{amsmath}    
\usepackage{amssymb}    

\begin{document}

\title{Detection of chemo-kinematical structures in Leo~I}
\titlerunning{Chemo-Kinematical Structures in Leo~I}
\author
{A.G. Alarc\'on Jara\inst{1}\thanks{E-mail: alexralarconj@udec.cl}, 
  M. Fellhauer\inst{1}, 
  J. Simon\inst{2}, 
  A. del Pino\inst{3,5}, 
  S.W. Fu\inst{4} \and
  S.T. Sohn\inst{5}}
\authorrunning{Alarc\'on et al.}
\institute{
  Departamento de Astronom\'ia, Universidad de Concepci\'on, Casilla
  160-C, Concepci\'on, Chile \and
  Observatories of the Carnegie Institution of Washington, 813
  Santa Barbara St., Pasadena, CA 91101, USA \and
  Centro de Estudios de F\'isica del Cosmos de Arag\'on
  (CEFCA), Unidad Asociada al CSIC, Plaza San Juan 1, 44001, Teruel,
  Spain \and
  Department of Physics and Astronomy, Pomona College,
  Claremont, CA 91711 \and
  Space Telescope Science Institute, Baltimore, MD 21218, USA} 

 \date{Received XXX; accepted XXX}

\label{firstpage}

\abstract{
  A variety of formation models for dwarf spheroidal (dSph) galaxies
  have been proposed in the literature, but  generally they  have not
  been quantitatively compared with observations.}
{
   We search for chemodynamical patterns in our observational data set
   and compare the results with mock galaxies consisting of pure
   random motions, and simulated dwarfs formed via the dissolving star
   cluster and tidal stirring models.}
{ 
   We made use of a new spectroscopic data set for the Milky Way dSph
  Leo~I, combining 288 stars observed with Magellan/IMACS and existing
  Keck/DEIMOS data, to provide velocity and metallicity measurements
  for 953 Leo~I member stars.  We used  a specially developed
  algorithm called {\sc Beacon} to detect chemo-kinematical patterns
  in the observed and simulated data.} 
{
  After analysing the Leo~I data, we report the detection of 14 candidate
  streams of stars that may have originated in disrupted star
  clusters.  The angular momentum vectors of these streams are
  randomly oriented, consistent with the lack of rotation in Leo~I.
  These results are consistent with the predictions of the dissolving
  cluster model.  In contrast, we find fewer candidate stream signals
  in mock data sets that lack coherent motions $\sim99$\% of the time.
  The chemodynamical analysis of the tidal stirring simulation
  produces streams that share a common orientation of their angular
  momenta, which is inconsistent with the Leo~I data.} 
{ 
  Even though it is very difficult to distinguish which of the
  detected streams are real and which are   only noise, we can
  be certain that there are more streams detected in the
  observational data of Leo~I than expected in pure random data.}


\maketitle

\section{Introduction}
\label{sec:intro}

The Local Group consists of about 80 galaxies discovered to date
\citep{McConnachie2012}, and this number is expected to grow with
observations of future telescopes and surveys.  Most of them are dwarf
galaxies, orbiting the two larger galaxies, the Milky Way (MW) and
Andromeda (M31).  Their classifications range from dwarf irregulars,
dwarf ellipticals, and compact ellipticals to dwarf spheroidal (dSph)
galaxies, the last composing the majority of the known dwarfs. 

Based on their high velocity dispersions and low luminosities, dSph
galaxies have high mass-to-light ratios and are believed to be the
most dark matter (DM) dominated stellar systems known
\citep{Mateo1998,Walker2009}.  They are among the oldest structures
and are by far the most numerous galaxies in the Universe; however, due
to their intrinsic faintness, the study of dSph galaxies has been very
difficult.  In the standard hierarchical galaxy formation models,
dwarf galaxies are the elemental systems in the Universe; larger
galaxies are formed from smaller objects like dwarf galaxies through
major and/or minor mergers \citep{Kauffmann1993,Cole1994}.  Thus, it
is important to study these galaxies to understand the formation and
evolution of normal sized galaxies.

DSph galaxies have a low stellar content and are poor in, or entirely devoid
of, gas.  The faint and ultra-faint population of dSph galaxies are
widely thought to be the smallest DM structures that contain stars
\citep{Mateo1998,Lokas2009,Walker2009}. The larger classical dSph 
galaxies are characterized by absolute magnitudes in the range $-13
\leq M_{\rm V} \leq -9$ \citep{Mateo1998,Belokurov2007}.  Their total
estimated masses, considering the stars and the DM halo is of the
order of $10^9$~M$_{\odot}$.  They exhibit high velocity
dispersions \citep[e.g.][]{Simon2007,Koch2009} in the range of $5-12$
km\,s$^{-1}$;  the velocity dispersion remains approximately constant
with distance from the centre of the galaxy
\citep{Kleyna2002,Kleyna2003,Munoz2005,Munoz2006,Simon2007,Walker2007}. 

Leo~I is one of the classical dSph galaxies orbiting the MW
detected by A.G. Wilson in 1950 \citep{Harrington1950} and described in 
\citet{Hodge1962}, among others.  It is one of the most distant MW
satellites.  At $257 \pm 13$~kpc \citep{Sohn2013} and with a
half-light radius of $244 \pm 2$~pc \citep{McConnachie2012}, it is
classified as a dSph galaxy \citep{McConnachie2012}, based on its
current morphology and lack of gas \citep{Knapp1978,Grcevich2009}.

Leo~I displays a very extended star formation history
\citep{Gallart1999a,Hernandez2000}, and is considered one of the 
youngest dSph galaxies in our Local Group in terms of average age
\citep{Lee1993,Weisz2014,Weisz2014b}.  Most of the  star-forming
activity in Leo~I happened between 7~Gyrs and as recently as 1~Gyr ago
\citep{Gallart1999a}.  Similar to others dwarf galaxies, Leo~I has a
low metallicity (of the order of ${\rm   [Fe/H]} = -1.43$~dex;
\citealt{Kirby2011}), contains no globular clusters, and is a purely
dispersion-supported system.  

\begin{figure*}
  \centering
  \includegraphics[width=\hsize]{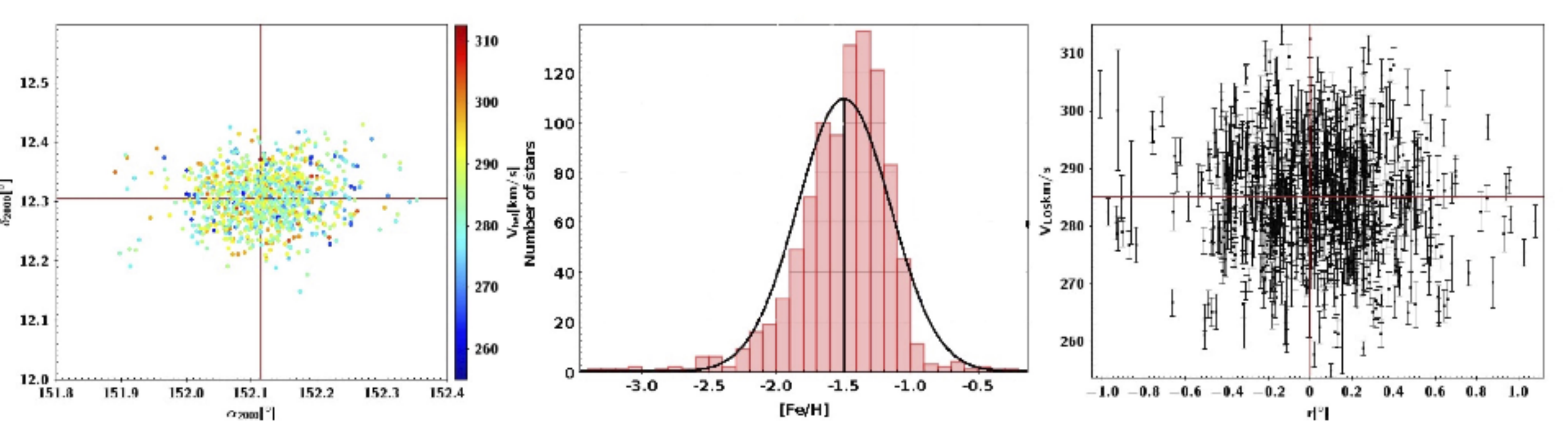}
  \caption{Illustration of our Leo~I data set.  The left panel shows
    the spatial distribution of the stars, colour-coded by velocity.
    The middle panel shows the metallicity distribution;   the mean
    metallicity for the galaxy is shown (black curve).  The right panel 
    shows the radial
    velocities as a function of right ascension, which closely
    corresponds to the major axis of the galaxy.}
   \label{fig:leodat}
\end{figure*}

Given its distance from the Milky Way and recent star formation
history, it provides an important tracer to explore the MW mass 
\citep{BoylanKolchin2013} and evolution models.  Metallicity studies
\citep{Bosler2007}, proper motion \citep{Gaia2018,Sohn2013}, and
radial velocity measurements \citep{Koch2009,Sohn2007} have
provided a significant amount of data on the galaxy, which has been
extensively analysed by various groups. 

Studies using statistical comparisons with simulations and direct
comparisons with Jeans spherical dynamical models often yield 
contradictory pictures. For example,   \citet{Mateo2008} found evidence for
an extended DM halo by observing a flat velocity dispersion profile
at large radii, and \citet{Lokas2008} found a DM profile similar to
the stellar profile.  Recently, \citet{Bustamante2021}
measured the central kinematics of Leo~I and find that the orbits of
the inner stars of the galaxy indicate the presence of a central black
hole of $3.3 \pm 2 \times 10^{6}$~M$_{\odot}$.  All this, combined
with the narrow stellar metallicity displayed by Leo~I, its shallow
negative metallicity gradient, and its positive age trend
\citep{Gullieuszik2009}, make Leo~I one of the most exciting dwarf
galaxies in the Local Group. 

Analysis of the proper motion of Leo~I indicate that this galaxy stopped
forming stars about one Gyr ago, probably due to the pericentric approach 
to the MW \citep{Sohn2013,Gaia2018}. 
Previous works have hinted at a turbulent past characterized by
alternating periods of intense star formation with quiescent intervals
\citep{Gallart1999a,Hernandez2000}. 

There are several models that attempt to explain the origin and
features of dSph galaxies by considering different mechanisms.  Some
of them are based on tidal and ram-pressure stripping
\citep{Gnedin1999,Mayer2007}.  In these models the dSph galaxies are
formed due to the interaction between a rotationally supported dwarf
irregular galaxy and a MW-sized host galaxy.  These models show that
dSph galaxies tend to appear near massive galaxies, but they do not
explain the presence of distant isolated dSph galaxies, such as
Tucana, Cetus, and Leo~T.  

The model proposed by \citet{DOnghia2009} considers a mechanism known
as resonant stripping, which can be used to explain the formation of
isolated dSph galaxies.  Basically, these objects are formed after
encounters between dwarf disc galaxies in a process driven by
gravitational resonances. 

According to the dissolving star cluster scenario
\citep{Assmann2013a,Assmann2013b,Alarcon2018}, a dSph galaxy is formed
by the fusion and dissolution of several star clusters (SCs) with low
star formation efficiency (SFE), formed within one DM halo.  This model
does not need gravitational interactions with other galaxies to
explain the formation of dSph galaxies, and can therefore can account for the
formation of isolated galaxies as well.

Several recent studies have been performed analysing the
chemo-kinematical patterns among the stars in multiple dwarf galaxies
from the Local Group, applying different methods
\citep{Cicuendez2018,delPino2017,delPino2017b,Lora2019}.  Using our
new extended data set, we are searching for similar chemo-kinematical
patterns in the Leo~I dSph galaxy.

In the next section we describe the spectroscopic data set used and
introduce the method {\sc Beacon} used to search for chemo-kinematical
patterns. We present the results from simulations and
observations in Section~3, and we discuss them in Section~4.

\section{Data and methods}
\label{sec:data}

\subsection{Observations}
\label{sec:spec}

For this project we combine spectroscopic data for Leo~I from three
different samples.  We used the samples from \citet{Sohn2007} and 
\citet{Kirby2010}, obtained with the Deep Imaging Multi-Object Spectrograph
(DEIMOS, \citealt{Faber2003}) on the Keck II telescope. These data sets
provide 749 stars with radial velocities and metallicities.  We also acquired 
our own data set, making use of the IMACS spectrograph on the Magellan
 Baade Telescope \citep{Dressler2006}. This data set adds the spectra of 288 red
 giants stars.  The combination of these spectroscopic samples provides 
kinematics and metallicities for a total of 953 Leo~I member stars, a factor of
$\sim 3$ larger than was used in previous Leo~I analyses.

\begin{table*}
  \caption{Clustering parameters: Input values for {\sc Beacon} used
    in this work.} 
  \label{tab:iniresn}
  \begin{tabular}{||c c c|| c c c c c c||}
    \hline
    \multicolumn{3}{l}{Galaxy parameters} &
    \multicolumn{5}{l}{Clustering parameters} \\ \hline
    Coordinates & Mean velocity & Distance & Standardization weights  &
    Maxima & $\epsilon_{\rm c}$  & Standardization &  Uniqueness   \\  
   RAS, DEC & [km/s]& [kpc]& RAS, DEC, $v_{\rm LOS}$, [Fe/H]& ratio &
   &method&method \\ \hline 
   [152.1146, 12.3059]  & 285.1994 &  254 & [1,1,1,2]  & 0.85 & 1/4
   &  St.dev. & any       \\ \hline  
   \end{tabular}
\end{table*}

Figure~\ref{fig:leodat} shows the 953 stars from the final sample used
in this work in three panels, with the star positions in the left
panel, colour coded according to their radial velocities.  The red lines show
the centre considered for this work with ($\alpha_{2000}$,
$\delta_{2000}$)=(152.1146, 12.3059) \citep{Munoz2018}.
  
The middle panel shows the metallicity distribution and a Gaussian fit
with a mean of [Fe/H]$ = -1.498$~dex and a standard deviation of
$\sigma_{Fe/H} =0.346$~dex.  Using the MCMC method we calculate a mean
[Fe/H]$ = -1.48 \pm 0.01$~dex with a standard deviation of
$\sigma_{Fe/H} = 0.27$. 

Finally, the right panel shows all measured $v_{\rm LOS}$ along the
$\alpha$-axis centred in $\alpha_{2000} = 152.1146$ showing a mean
heliocentric velocity of the dSph galaxy, calculated with the MCMC
method, of $v_{\rm hel} = 285.2 \pm 0.3$~km\;s$^{-1}$.  We
measure a velocity dispersion of $\sigma = 9.2 \pm 0.2$~km\;s$^{-1}$.
With this we calculate the mass contained within the half-light radius
using the formula from \citet{Wolf2010}, valid for
dispersion-supported stellar systems in dynamical equilibrium.  Using
the half-light radius $R_{1/2} = 244 \pm 2$~pc measured by
\citet{Munoz2018}, following a Sersic profile, gives us an enclosed
dynamical mass of $M_{1/2} = 1.9 \pm 0.1 \times 10^7$~M$_\odot$
which is in agreement with previous calculations. 

\subsection{{\sc Beacon}: a software that finds chemo-kinematical patterns} 
\label{sec:beacon}

The {\sc Beacon} software package was developed by \citet{delPino2017}
to search for rotation patterns in resolved stellar populations.  The
code is publicly
available,\footnote{\url{https://github.com/AndresdPM/BEACON}} and is free
to use and modify.  It is written in Python, and was created to check the
results of \citet{delPino2015}, who show that the Fornax dSph galaxy
contains different stellar populations with different angular momenta.

{\sc Beacon} is designed to find groups of stars with similar
positions, velocities, and metallicities or chemical compositions.
Based on the {\sc Optics} clustering algorithm \citep{Ankerst1999} it
searches in a given data set for groups of stars with similar chemical
composition and velocities, and if the number of stars in the group is
bigger than a minimum cluster size (MCS), the code classifies them as
a one-side stream (OSS).  If on the opposite side of the centre of
mass (CM) there isi a stream with similar chemical composition, but with
opposite velocity, then the two groups are classified as a both-side
stream (BSS) or circular stream.  The full description and clustering process 
is described in \citet{delPino2017}, their section 3.

Apart from the observational data, we needed some galaxy parameters and
clustering parameters in order to create a reference frame centred on
the CM of the galaxy and to control the clustering process. 
 The galaxy parameters are the coordinates of the CM and the $v_{\rm
  hel}$ of the CM, which are necessary to detect rotation patterns in
the galaxy.  The clustering parameters, on the other hand are a
collection of parameters defining the clustering criteria: the standardization 
method and the standardization weight, the uniqueness of solution, the maxima 
ratio, and the inimum cluster size (MCS). 

The standardization method and the standardization weight are used
for the standardization of the state vector, which affect the
relative importance of each coordinate during the clustering. 
Depending on the uniqueness of solution, {\sc Beacon} can merge clusters that 
share stars into one cluster, thus avoiding duplicate stars, i.e.\ considering 
clusters different if they differ in at least one star. 
The maxima ratio is the reachable distance ratio; the higher the
maxima ratio is, the more sensitivity {\sc Beacon} has, making it
prone to spurious detections. 
The minimum cluster size (MCS) is the minimum number of stars that a
cluster should have to be considered  a stream during the
clustering process.  

All these parameters have to be tuned for the specific galaxy under
study and/or the scientific case, and depend  on the number
of sampled stars and on the completeness of the sample and its spatial
coverage.  We use the parameters given in Table~\ref{tab:iniresn} and
use MCS as a parameter we vary.  We give a stronger weight to the
metallicity in our study as we want to compare the observations
mainly to the dissolving star cluster model in which we expect the
stars in one   cluster to have a similar metallicity.  In addition, the
weight on the two coordinates is higher, as it is  in studies where the
authors are searching for an overall rotation pattern.

In previous studies the weights on positions and metallicities were
lower as the authors were searching for an overall rotation pattern in the
data.  If we expect a rather flat rotation curve (as seen in dSph
galaxies in the velocity dispersion versus radius results), it is
possible to enhance the signal by placing less weight on the
positions.  Furthermore, expecting that the rotation is being
supported by stars from different populations or at least different
star-forming events, placing a weight on the metallicities would be
counterproductive and make the detection much more difficult. 

In contrast, when trying to find evidence for the dissolving star
cluster scenario (DSCS), this choice of parameters will not work at
all and will lead to a much higher rate of false detections.  In the
DSCS we expect each possible stream to stem from the dissolution of a
small star cluster or association, formed inside the central region of
the dwarf galactic DM halo; in other words,  the stars should  not only have 
similar velocities, but should also be coherent in space and
have the same metallicity.  Such streams are meanwhile found in our MW
in the recent GAIA data catalogues.  Only when placing weights on
positions and metallicity do such structures become visible.  On the other
hand, our choice of weights does not inhibit the detection of a grand 
design rotation, it is just  more difficult to find  within the
noise.  The destroyed dwarf disc models we introduce at the end of our
manuscript show that we are in fact able to recover  an overall
rotation pattern as well.

\subsection{Mock data}
\label{sec:mock}

\begin{figure}
  \includegraphics[width=\hsize]{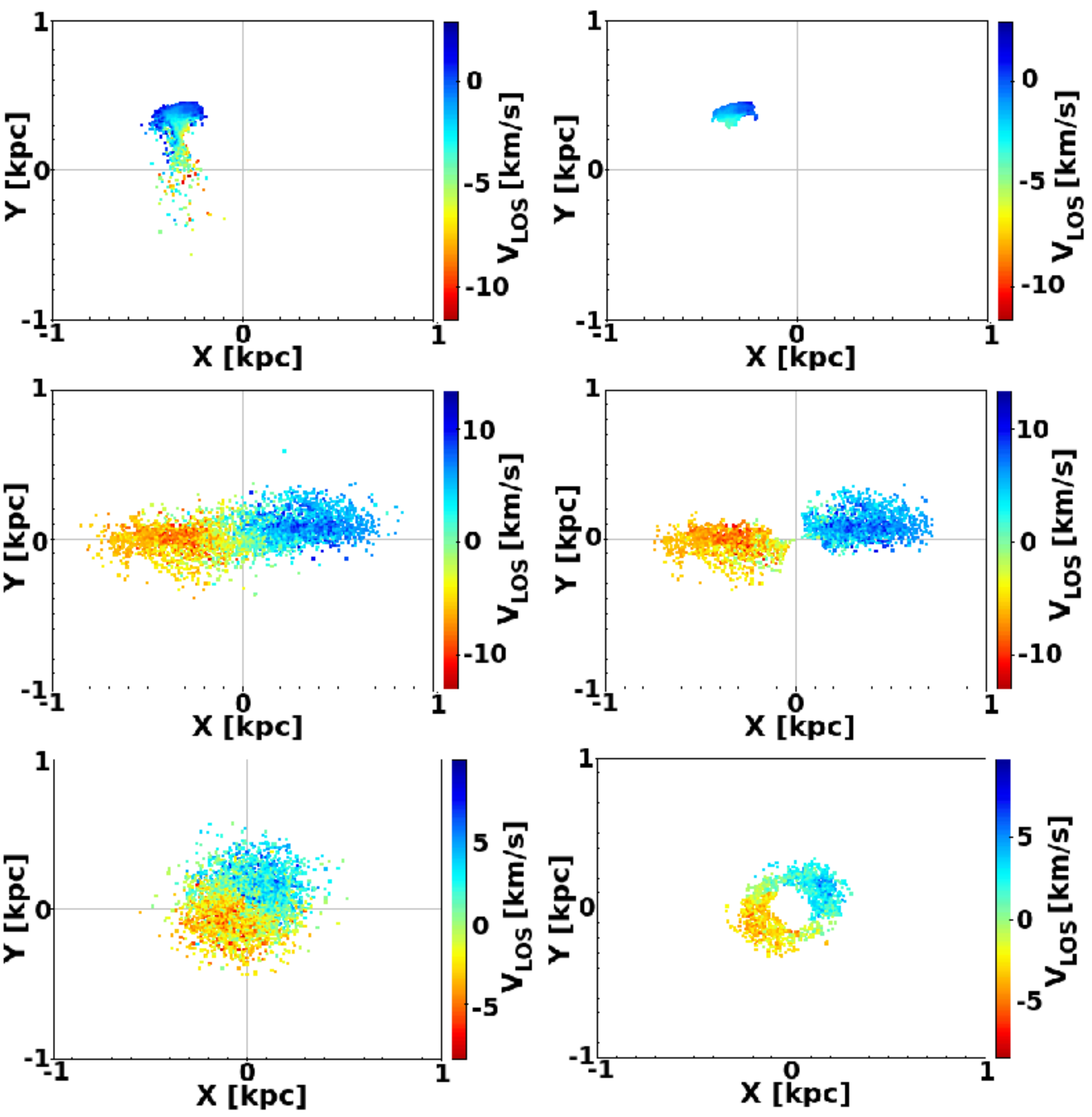}
  \caption{Shapes of streams produced by dissolving star clusters in 
    the dissolving star cluster model.  The panels in the left column
    show all particles of a single star cluster in the chosen sample,
    while the right column includes only the stars of the same star
    cluster that are recovered and flagged as a stream by {\sc
      Beacon}.  The top row shows a recently dissolved cluster
    recovered by {\sc Beacon} as a one-side stream.  The middle
    row illustrates a cluster orbiting mainly in the x-z plane. The
    bottom row shows a cluster orbiting mainly in the x-y plane.  {\sc
      Beacon}   recovers the main feature of the one-side stream
    of the recently dissolved cluster.  It also recovers the both-side
    streams of the two other clusters independent of the orbital
    orientation.  It is clear that {\sc Beacon} has problems 
    recovering the stars in the central region, which do not have large
    velocity differences.  This behaviour is caused by the relatively
    large weights for the spatial coordinates.}
   \label{fig:formastream}
\end{figure}

To be able to compare our observational results with purely
randomized mock data, we constructed two types of data sets that
have no streaming motion.  

The first type of data set represents a
Plummer sphere with a Plummer radius equal to the half-light radius
of Leo~I.  The Plummer mass is chosen to obtain a  velocity
dispersion similar to that of  Leo~1 (i.e.\ the sphere is in virial 
equilibrium), but
has no sign of rotation or any streaming motion.  From the
distribution function we construct 1000 samples of 1000 radial
velocities.  To these sets of `exact' data we add random errors
according to the observational uncertainties of our Leo~I data.

As a second type we constructed data sets based on a similar Plummer
distribution, but taking the shape of Leo~1 into account (i.e.\ the
same eccentricity as Leo~1).  The velocities are corrected as if a DM
halo is present, as implied by the observations, but having no rotation
or streaming motions.  This model is not in equilibrium by itself,
but is the   closest mock representation of Leo~1 without actual
rotation or streaming motion.  Again, we draw 1000 samples of
1000 radial velocities from the resulting distribution function and
convolve the data sets with the observational errors.

This procedure is statistically equivalent to analysing 1000
data sets of 1000 different objects (i.e.\ dSph galaxies) with
similar parameters.

\subsection{Simulations}
\label{sec:simul}

\begin{figure}
  \includegraphics[width=\hsize]{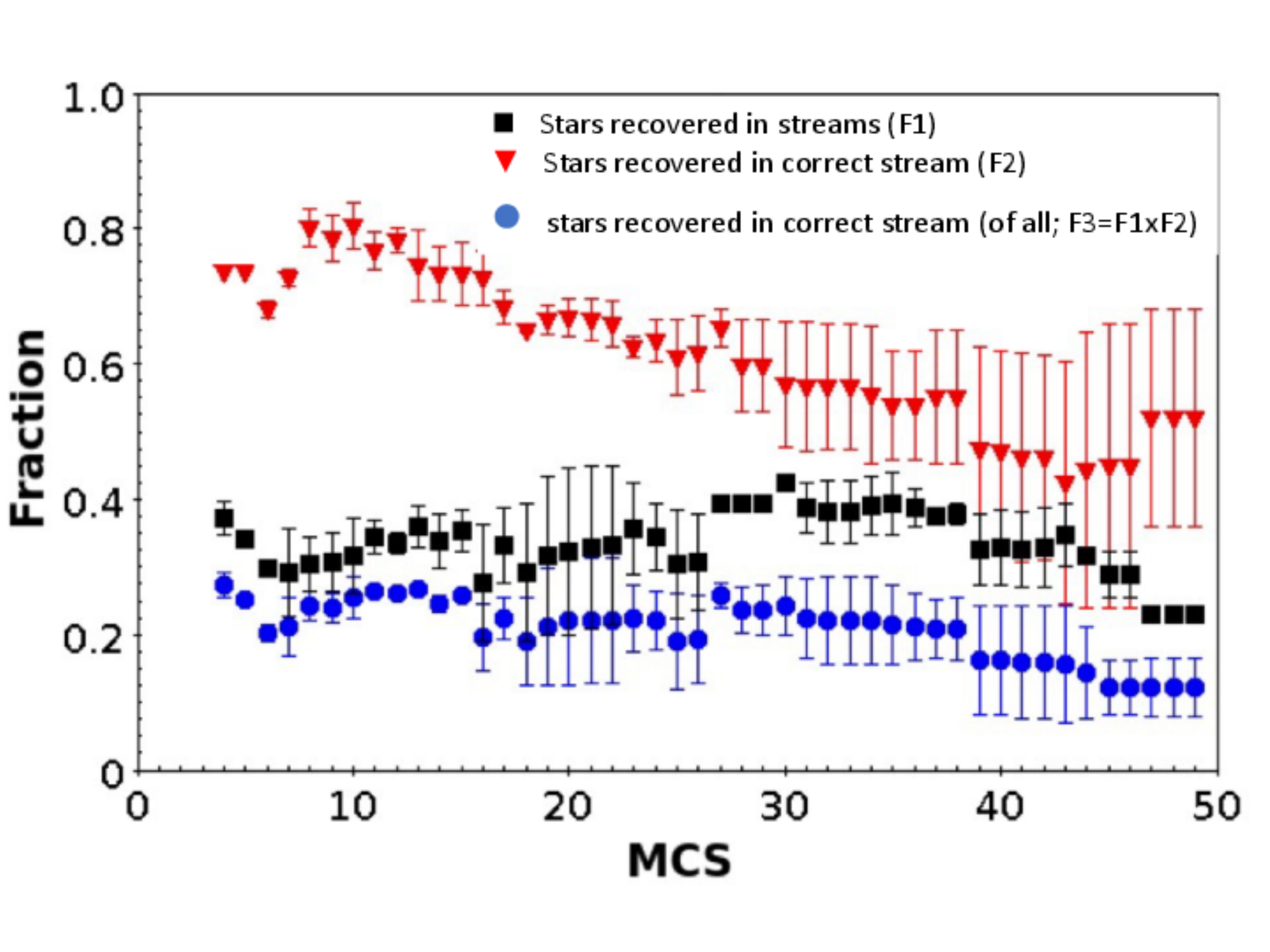}
  \caption{Fractions of recovered stars as functions of the clustering 
    parameter MCS (all other parameters are kept constant, as shown in
    Tab. \ref{tab:parstrleo}).  We show as F1 the fraction of stars
    assigned to streams as black squares.  As F2 we show the fraction
    of correctly assigned stars in relation to the recovered
    stars.  Finally, as F3 we show the fraction of correctly recovered
    stars compared to all stars as blue circles.  We see that {\sc
      Beacon} is indeed able to recover about 20 to 25\% of
    stars as belonging to streams correctly, even though the data is
    degraded by  errors similar to those in the observational data.} 
   \label{fig:eficiencia}
\end{figure}

In this study we focus on the dissolving star cluster model, as
described in \citet{Assmann2013a}, \citet{Assmann2013b}, and 
\citet{Alarcon2018}.  We use
simulations with $16$~star clusters (SCs), placed in a cusped DM halo 
following a Navarro Frenk \& White (NFW) profile 
\citep{Navarro1997}
with a scale length of $r_{\rm s} = 1$~kpc,  modelled with 1,000,000
particles and an enclosed mass at $500$~pc of $M_{500} =
10^{7}$~M$_\odot$.  For the $16$~star clusters, we adopt a SFE$ =
20$\,\%, amounting to a final total mass of stars $M_{\rm stars} = 4.5
\times 10^{5}$~M$_\odot$.  Each star cluster itself is modelled using
100,000 particles, following a Plummer distribution with a Plummer
radius of $r_{\rm sc} = 4$~pc.

The star clusters are inserted into the simulation according to a
constant star formation history, as described in \citet{Alarcon2018} (i.e.\ forming one star cluster every 625~Myr).  For simplicity, we
assign a metallicity to each star cluster following a linear age-to-metallicity ratio given by 
\begin{eqnarray}
  \label{eq:metal}
  [{\rm Fe/H}] & = & -0.0004 \cdot {\rm Age[Myr]} + 0.3267.
\end{eqnarray}

The initial positions and velocities of the star clusters are randomly
drawn from a Plummer distribution to ensure that more clusters are
inserted in the central area than in the outskirts of the DM halo.
The velocities are corrected for the presence of the DM halo to keep
the orbit of the SCs the same as without the DM halo.  

The simulations are carried out using the particle-mesh code {\sc
  Superbox} \citep{Fellhauer2000}.  For more details about the initial
conditions and the simulations carried out, we refer to the previous
study of \citet{Alarcon2018}.
After the simulations we add randomly drawn errors to the exact values
given by the final simulation data (i.e.\ to the radial velocities and
metallicities of our `stars') according to the mean errors of our Leo~I
sample which are $\pm 2.5$~km\,s$^{-1}$ and $\pm 0.2$~dex.  

In these models the star clusters orbit the centre of the dark matter
halo while they dissolve leaving streams of stars that conserve their 
kinematic properties.  This is the main prediction of this model, which 
could be corroborated with spectroscopic observations.  Examples of
these rotating patterns are shown in the left panels of
Fig.~\ref{fig:formastream}, which show their positions and velocities.
To produce Fig.~\ref{fig:formastream} a larger sample of particles
from each SC is used (i.e.\ 7000 per SC) to better present the
qualitative behaviour than in the subsequent analysis and
to ease comparison of the different data sets.

The first panel shows a young star cluster that is not entirely
dissolved, but is in the process of dissolution;  after a few
hundred Myr more its stars will be dispersed around the galaxy.
The second and third panels show the stars for clusters that spread
their stars completely around the dark matter halo, it is clear to see
the streaming motion in these cases.  Most of the stars with positive
velocities are on one side of the dark matter halo;  on the opposite
side we see the stars with negative velocities.

The right panels show the stars that {\sc Beacon} is able to recover
for each stream.  We can see that it has difficulty  detecting the stars
in the centre as they have relative velocities close to zero and
their positions are far from the stars that have similar velocities.
The lack of sensitivity in the centre can be corrected by setting a
smaller weight for the radial coordinate of the stars.  This would
allow {\sc Beacon} to find more radially distributed groups, but
it could also negatively impact the purity of the resulting detected
clusters.  

To compare the data sets from the simulations with our observational data, we
took 62 random particles from each star cluster, which gave  us 992
particles in total for each set. This means that each simulated data set has an
 amount of stars that is similar to  our observational data set. 

\section{Results}
\label{sec:res}

\subsection{Simulations}
\label{sec:simres}

\begin{figure*}
  \includegraphics[width=\hsize]{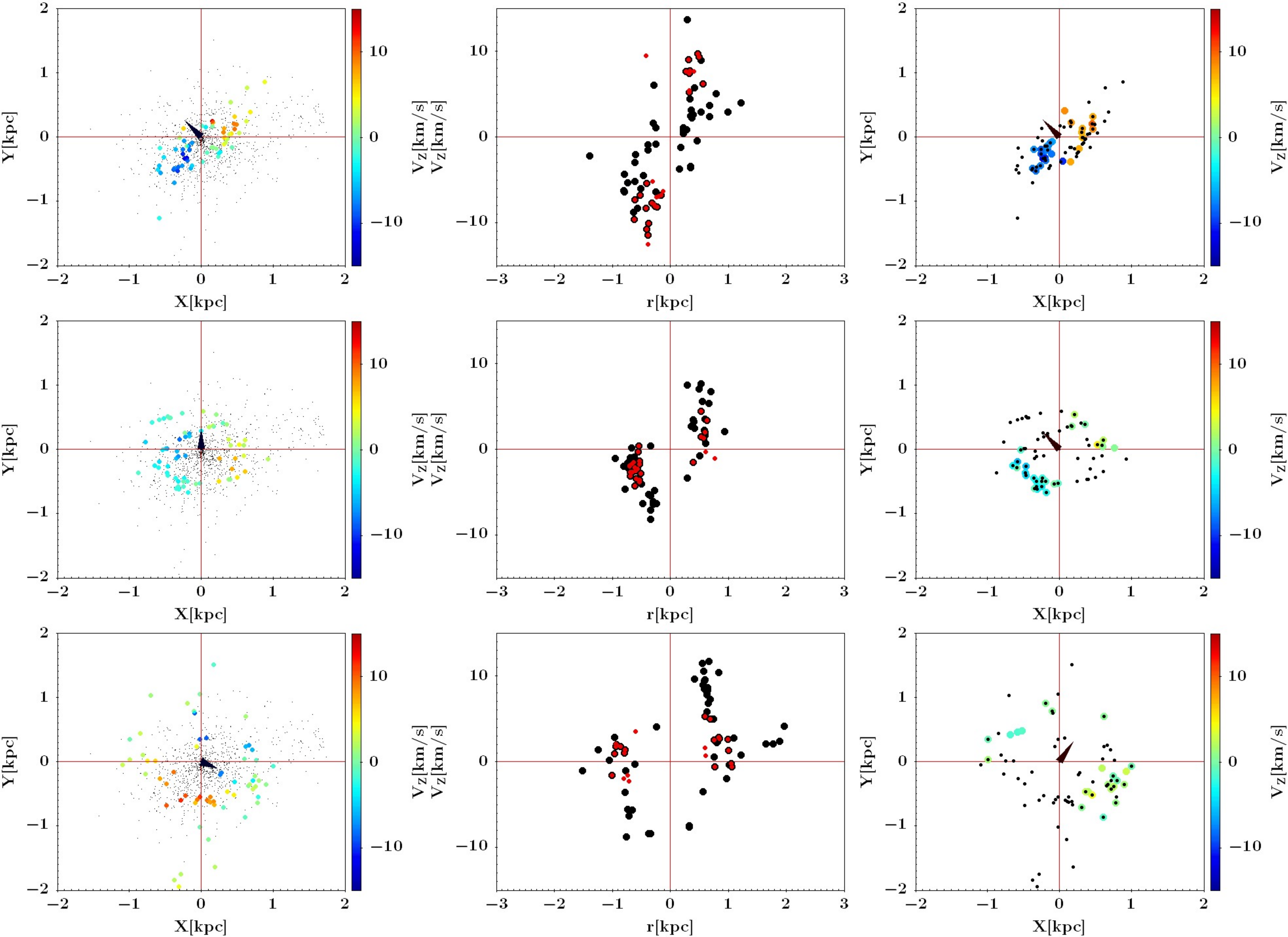}
  \caption{An example of how {\sc Beacon} can extract substructures hidden
    in a large data set with 992 particles.  Here we apply {\sc Beacon} to a
    simulation of dissolving star clusters with 16 star clusters and a
    constant star formation.  The left panel shows the position of the
    stars and the velocities along the Z-axis are shown as colours for
    the stars that were formed in the same star cluster. The black
    triangles show the projected angular momentum direction and black
    dots are all other particles from the simulation.  Middle panels
    show the velocities along the line of sight taking as a new Y-axis
    the projected angular momentum direction, black dots are the
    values for the stars which belong to the same star cluster and red
    dots are the stars recovered by {\sc Beacon}.  In the right panels we
    show the particles recovered by {\sc Beacon} in colour and in black the
    stars which belong to the same star cluster.  In these examples
    {\sc Beacon} was able to recover most of the stars which show a strong
    rotation signal, with a few false detected stars which do not
    belong to the star cluster.} 
  \label{fig:bcsim}
\end{figure*}

It is a very difficult task to correctly recover all stars of all
streams,  especially once we add errors similar to the observational
data.  Instead of sharp metallicity peaks, we have smeared out
distributions which might even overlap.  In the same spatial region we
might have stars from different dissolved star clusters but with very
similar radial velocities.  To understand better the efficiency of
{\sc Beacon} we define three parameters: F1, F2, and F3.

F1 is  the fraction of recovered stars. We count all that stars that
{\sc Beacon}   attributes to the different streams it finds.  We
divide this number by the total number of stars.
F2 is the fraction of correct stars (recovered).  We only count
the stars that  {\sc Beacon} has assigned to the correct stream, and
divide this number by all recovered stars.  
F3 is  the fraction of correct stars (all).  We count all correctly assigned
stars, but now divided by the total number of stars.
As in our simulations, all stars of the dwarf galaxy originate in
streams; the division by all stars also implies the ratio of stars
with respect to all stars originating in streams as well.

In Fig.~\ref{fig:eficiencia} we plot these three fractions as a
function of the clustering parameter MCS.  All other values are kept
constant at the values given in Table~\ref{tab:iniresn}.  We take the
data sets from the three simulations of the dissolving star cluster model.
This  gives results with their respective error margins.

The black squares show the fraction of stars that are assigned to
streams.  This fraction is  constant at 40~\% as long as MCS is
well below our sample size of 62~stars from each cluster.  As soon as
MCS  gets closer to our sample size, this value drops.

We can see that this is not the full story  if we look at the red
triangles, which record the fraction of correctly assigned stars.
Here we see a fraction of 70 to 80\% of stars     assigned to
streams that   actually belong to that stream, especially at MCS values
below 20.  Once we raise this parameter further this fraction drops to
about only half of the recovered stars.

Finally, and most importantly, the blue dots give us the
fraction of stars that are   correctly assigned to their streams of all
the stars in the sample.  This fraction is of the order of 20 to 25\%
for small MCS values and drops below 20\% if MCS gets closer to the
sample size.

This in turn means that even in simulations where we should be able to
recover all possible streams, once we restrict ourselves to an overall
sample of about 1000 stars and adding observational errors to our
data, we can only expect to recover about one-fourth of the stars correctly.
This could mean, in the worst case scenario,  that  
we might be missing    the detection of three-fourths of the streams actually
present in the data.  This is not seen in our samples based on
simulated streams, where we detect about 70\% of the
actual streams (see below).
These values give us confidence that we can use {\sc Beacon} on
an observational data set and that we will obtain a sufficiently good
answer; in other words,  that we are able to see about a quarter of the stars
assigned to their correct streams if there are any.

In Figure~\ref{fig:bcsim} we show   some examples of recovered
streams by {\sc Beacon} for one of the simulations using ${\rm MCS} =
10$.  The left panels show the positions of the stars with a  colour bar for
the velocities, the coloured stars are the stars that belong to the
same star cluster, and the black dots are the rest of the stars in the
simulated galaxy.  The black arrows show the direction of the projected
angular momentum.  It is clearly   that the stars from the same
dissolved star cluster follow similar orbits since the stars with
positive velocities are located at one side of the centre of the
galaxy, and on the opposite side are the stars with negative
velocities.  The middle panels show on the $x$-axis the distance to the
centre and on the $y$-axis $v_{z}$ the velocities.  Black dots are the stars
of the dissolved star cluster and red dots are the stars that {\sc
  Beacon} recovers as a stream.  We can see that {\sc Beacon} recovers
a large part of the stars from the sample, but a few of them were
missed   or misclassified. In addition,  some stars from other star
clusters are  misclassified as members of a particular stream.  In the
right panels we show the particles recovered by {\sc Beacon} 
and  the stars that belong to the same star cluster.  In
these examples {\sc Beacon} was able to recover most of the stars
which show a strong rotation signal, with a few false detected stars
that do not belong to the star cluster.   We can see that there are a
small mismatches between the projected angular momentum recovered and
the real value, due to the missed   stars. 

\begin{figure}
  \includegraphics[width=\hsize]{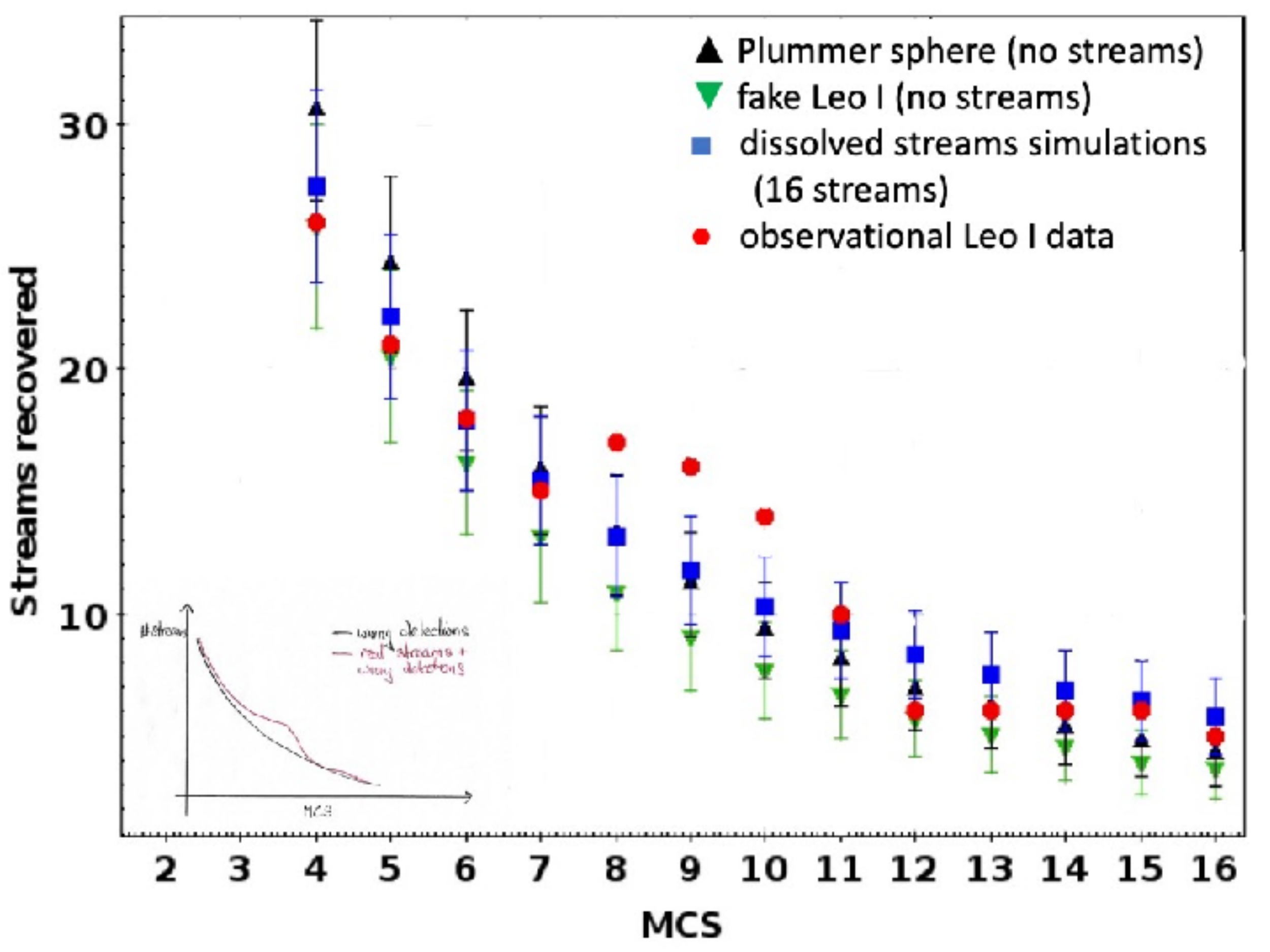}
  \caption{Number of streams recovered with {\sc Beacon} as a function
    of MCS. The blue rectangles are mean numbers taken from   three simulations
    of the dissolving star cluster model consisting of 16 streams.
    The black triangles (up pointing) are for a Plummer sphere distribution
    with no rotation patterns (i.e.\ no streams); the green triangles
(down    pointing)  are Leo~I fake data, also with no streams.  The red
    circles are the results of {\sc Beacon} applied to the
    observational data set of the Leo~I dSph.  We use 2 as the value for
    the weight of the metallicity, with a sensitivity of 0.85 and a change in
    the minimum cluster size MCS.  The   inset shows
    qualitatively how a few real streams should appear in a `sea' of
    false detections.}
   \label{fig:streamspleosim}
\end{figure}

\subsection{Comparison with mock data}
\label{sec:compmock}

As a next step we want to check how significant our detection of
streams is in comparison to mock data, which by design contains no
streaming motion whatsoever.

We construct samples of stars from our simulations and from the
mock data of non-rotating Plummer spheres and the fake Leo~I data.
The size of the samples is the same as the number of observed stars we
have.  To the exact data we add the observational errors   determined
for our Leo~I data set.  Now {\sc Beacon} is applied, keeping  the
parameters shown in Table~\ref{tab:iniresn}   constant and varying
the sensitivity by varying the parameter MCS.  The results are shown
in Fig.~\ref{fig:streamspleosim}. 

The black and green triangles show the detected streams in the mock
data, which by construction does not contain any streams.  We can see
clearly that the number of false detections decreases exponentially
with decreasing the sensitivity (i.e.\ requiring more stars for the
stream detection; increasing MCS).  Error bars to these values are
possible, as we construct 1000 samples of mock data. 

The blue squares show the number of detected streams in our dissolving
star cluster simulations, consisting of 16 different star clusters.
We use three different samples from three different simulations to be
able to place error bars on the measured values. 

If we  used the exact data of one sample, which would allow the
detection of all streams, we would expect a curve in which for small 
MCS the real detections would be surpassed by the noise of false
detections.  Once the false detections fall below the number of
real detectable streams, the data points should show an almost
constant value as function of MCS (i.e.\ the real streams plus some
small contribution of the noise).  Once MCS reaches the sample
size of the stars belonging to a stream, the detection signal should
go rapidly back to the noise level (see  inset of
Fig.~\ref{fig:streamspleosim}). 

As we are using multiple samples from different simulations, we cannot
expect such a curve.  Instead, we have to deal with a super-position of
a multitude of such curves, smearing out this clear signal.  Even so,
  we see  that as soon as we require an MCS $> 8$, the blue
symbols are always above the curves of false detections.  The
  blue symbols should fall back down within the noise level once we reach
  MCS values close to about 20\% of the actual sample size (i.e. stars
drawn from that particular dissolved cluster) as we expect only to
recover about 20-25~\% of the stars from a stream correctly (see
Fig.~\ref{fig:eficiencia}).
The red circles are based on the observational data of Leo~I which we
 talk about in the next section.

Unluckily, the small sample size of fewer than a thousand stars
combined with the observational uncertainties  leads to
overlapping error bars;  the real detections of streams are not
above the noise level by a statistically significant amount.

\subsection{Observational results}
\label{sec:obsres}

\begin{figure*}
  \includegraphics[width=\hsize]{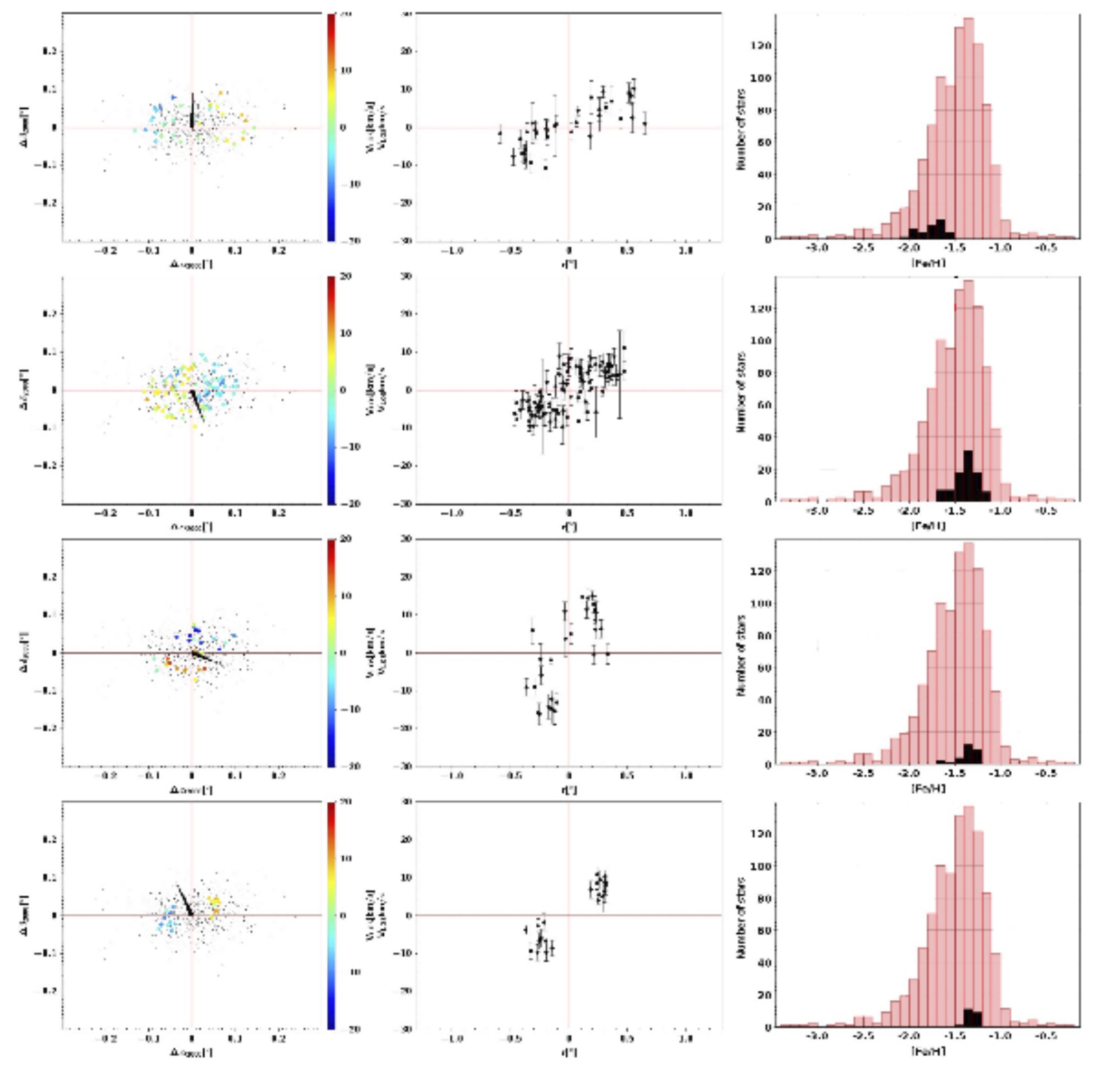}
  \caption{Examples of stream motions recovered by {\sc Beacon} in the 
    observational data of Leo~I.  Left panels: Positions of all
    observed stars as small black points.  The coloured points are the
    stars {\sc Beacon} joins to a stream detection;   the colour
    represents the radial velocity measured for each star.  The
    direction of the projected angular momentum of this stream is
    shown as a black arrow.  Middle panels: Radial velocity of
    the stars belonging to the stream as a function of their distance
    to the centre of the dwarf galaxy.  Right panels: Metallicity distribution of the whole sample (red) and  the
    metallicity distribution of the stars from the stream recovered by
    {\sc Beacon} (black).}
   \label{fig:streamsleo}
\end{figure*}

\begin{table}
  \caption{Parameters from the BSS recovered by {\sc Beacon} in Leo~I.
Column 1 shows the number assigned by {\sc Beacon}, Col. 2   the mean metallicity of the BSS and its standard
    deviation, Cols. 3 and 4   the angle (with respect
    to the major axis) and the modulus of the projected angular
    momentum, Col. 5 the number of stars in each BSS.} 
  \label{tab:parstrleo}
  \begin{tabular}{|c|c|c|c|c|c|}
    \hline
    Pop  & [Fe/H]  & Angle & Modulus & Number   \\ 
    &         &   [$^{\circ}$] & [$\times10^3$ pc$^2$/s]  & of stars  \\ \hline
    1 & -1.89 $\pm$ 0.26 & 263.5 & 1.79 & 21 \\
    2 & -1.71 $\pm$ 0.13 & 89.4 & 1.57 & 35 \\ 
    3 & -1.68 $\pm$ 0.07 & 285.0 & 0.50 & 37 \\
    4 & -1.63 $\pm$ 0.30 & 82.6 & 1.04 & 31 \\ 
    5 & -1.43 $\pm$ 0.13 & 43.1 & 0.13 & 20 \\
    6 & -1.41 $\pm$ 0.11 & 49.3 & 0.51 & 19 \\
    7 & -1.38 $\pm$ 0.13 & 288.0 & 0.90 & 88 \\
    8 & -1.35 $\pm$ 0.09 & 105.0 & 1.46 & 12 \\ 
    9 & -1.36 $\pm$ 0.10 & 336.9 & 0.65 & 28 \\
    10 & -1.32 $\pm$ 0.05 & 112.1 & 1.67 & 21 \\
    11 & -1.31 $\pm$ 0.06 & 268.3 & 0.92 &  22 \\
    12 & -1.24 $\pm$ 0.04 & 144.9 & 0.03 &  24 \\
    13 & -1.16 $\pm$ 0.05 & 212.4 & 0.12 & 35 \\
    14 & -0.74 $\pm$ 0.29 & 232.5 & 0.22 & 13 \\ \hline
  \end{tabular}
\end{table}

In Fig.~\ref{fig:streamspleosim} the black and green triangles
represented the mock data.  In an ideal world we would like the
response of {\sc Beacon} to be zero.  This is not the case, and if we
reduce the sensitivity of {\sc Beacon} such that we get an almost zero 
response in the no-stream data sets, we also get a similar response
for the simulation data, where streams are clearly present (blue
squares).  For very high sensitivities the response of {\sc
  Beacon} gives us more streams than  are actually in the data.

To   analyse the observational data of Leo~I we require that {\sc
  Beacon} does not recover more streams than are actually present for the
dissolving star cluster data (i.e.\ 16 streams for the blue
symbols).  Furthermore, we want the false detection (green and black)
to be below the number of streams detected in the real observational
data (red dots). These requirements are matched using a MCS range
between 8 and 11.

Furthermore, we note that the red data points follow that  pattern that we
  predict theoretically in Sect.~\ref{sec:compmock}.  This could
be a clear sign that there is actually detectable streaming motion
present in the observational data of Leo~I.

 \begin{figure*}
  \includegraphics[width=\hsize]{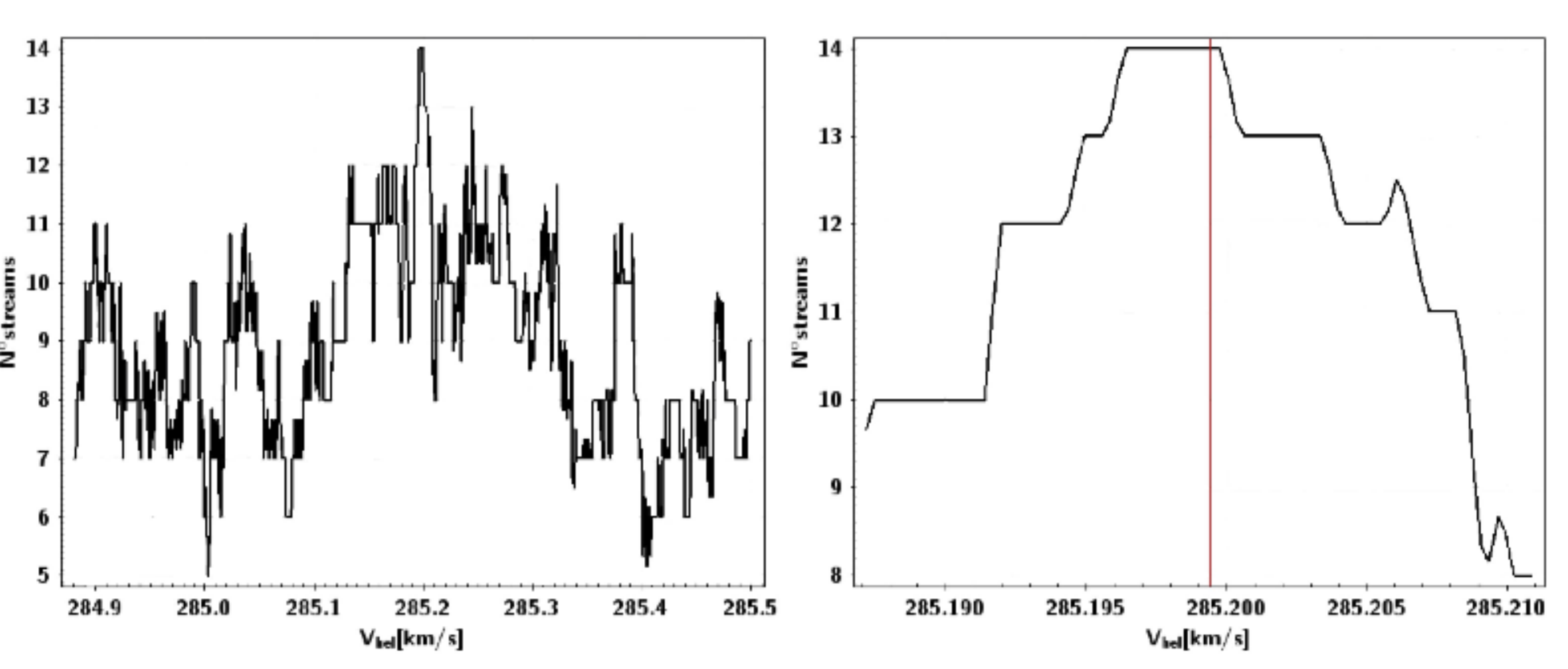}
  \caption{Number of recovered streams using ${\rm MCS} = 10$ vs\ the mean
    heliocentric velocity of Leo~I $v_{\rm hel}$.  Left panel: We apply 
    {\sc Beacon} varying the $v_{\rm hel} = 285.2$ between $\pm 0.31$ with 
    steps of $0.0001$~km\,s$^{-1}$.  The maximum number of streams is 
    recovered between $285.196442 < V_{\rm hel} < 285.199738$~km\,s$^{-1}$.
    Right panel: Magnified image of the central region shown in the left 
    panel.  The adopted $V_{\rm hel} = 285.1994$~km\,s$^{-1}$ in our study is
    shown as a red vertical line.} 
   \label{fig:vhelvsNstream}
\end{figure*}

For the further analysis we   apply a MCS parameter of 10.  With
this value we find $14$ detections of probable streams with {\sc Beacon}.
At ${\rm MCS} = 10$ we find $11 \pm 2$ streams out of $16$ in our
simulations of the dissolving star cluster model, and we can calculate
the probability of  finding $14$ or more streams in the mock data, which
is as low as $1.7$~\% in the case of the Plummer sphere data and just
$0.4$~\% in the case of Leo~I mock data, making it a statistically
significant detection.  

In Fig.~\ref{fig:streamsleo} we show some examples of the streams
detected in the observational sample of Leo~I.  The left panels show
the positions of all measured stars as black dots;   the stars
recovered as a stream are coloured according to their radial velocities.
The black arrows show the direction of the projected angular
momentum.  In the middle panels the radial velocities as function of
the distance to the centre are shown, and in the right panels 
the metallicity distribution of all measured stars in red and the
metallicities of the recovered stars of that particular stream as black
histograms.  
The fact that $\approx 40$~\% of the stars were classified in a stream
is in agreement with the values recovered by {\sc Beacon} in a
simulation with 16 star clusters using the same parameters (see
Fig.~\ref{fig:eficiencia}, where the  black boxes indicate the fraction
of stars recovered in a simulation of the dissolving star cluster
model:  $\approx 35\%$).

No OSS is detected by {\sc Beacon} in Leo~I.  The sample of stars used
in this work consider just RGB stars that are older than 2~Gyr, so
they would have enough time to be dispersed around the galaxy to show
up as a BSS. 

After applying {\sc Beacon} to our catalogue with input parameters
shown in Table~\ref{tab:iniresn} a total of 406 stars were classified
into 14 possible circular streams.  The other stars were considered
as stars that do not follow any strong chemo-kinematical pattern.
Some of these stars may belong to rotating streams and may have been
misclassified, as in the simulations.  Some stars show nearly zero
$v_{\rm LOS}$, making their classification into BSSs difficult.  Some
others may simply belong to poorly sampled populations that have not
fulfilled the minimum requirements to be considered as a group.  Given
the fraction of correctly recovered stars in the simulated data (F3),
these results are consistent with most or all of the Leo~I stellar
population originating in clusters that have now   dissolved,
spreading their stars inside the DM halo of the dSph, but still
following orbits similar to their original SC.

Table~\ref{tab:parstrleo} shows the properties of the 14 streams
recovered by {\sc Beacon}.
The projected angular momentum can be derived for each BSS,
$ {\bf L} = \sum_{i=1}^{i=j} {\bf r}_{i} \times m_i {\bf v}_{i,{\rm
    LOS}}$, where $m_i$ is the mass of the $i$-{th} star and $j$ is
the number of stars in the BSS.  A priori, we do not have information
about the homogeneity of our sample in terms of metallicity.  A
sensible way to avoid sampling effect is to normalize the ${\bf L}$
vector of every group to the number of stars in the group.  This is
roughly the same as expressing the angular momentum per unit of
stellar mass and provides a measure of the mean momentum ${\bf p}$ of
the stars forming a BSS.  In the following we use only quantities
normalized in this way.  

The question arises of  why, in Table~\ref{tab:parstrleo}, all streams
show member counts of 19 and more, while the red data points of
recovered streams as function of MCS disappears into the noise already
at values equal to 12.  This is due to the fact that {\sc Beacon}
after a first detection (using MCS) is searching for more stars that
could belong to the stream, and is also grouping similar streams
together.  Finally, once a OSS is detected, it searches on the other
side of the galaxy for a counterpart to combine  into a BSS.  As
stated above, all recovered streams in the Leo~I data are BSSs.

\subsection{Some caveats}
\label{sec:worries}

As a final test we vary the mean heliocentric velocity $v_{\rm hel}$
of Leo~I in the application of {\sc Beacon} on our observational
data.  As shown before, streams with velocities almost perpendicular
to the line of sight,  as well as stars in a stream close to the centre
of the dwarf with relative velocities close to zero, are
undetectable by {\sc Beacon}.  So we expect the detection of some
streams to be highly dependent on the correct choice of the mean
velocity of Leo~I.  The results of this test are shown in
Fig.~\ref{fig:vhelvsNstream}.

Both panels of the figure show the number of recovered streams as a
function of the mean heliocentric velocity, which we have used in {\sc
  Beacon} as a fixed parameter so far.  The right panel is a zoomed-in image of
the central part of the left panel.  We see strong variations in the
number of recovered streams when changing the adopted velocity by just
a few metres per second;   the maximum number is at the determined
systemic velocity of Leo~I.

\begin{figure*}
  \includegraphics[width=\hsize]{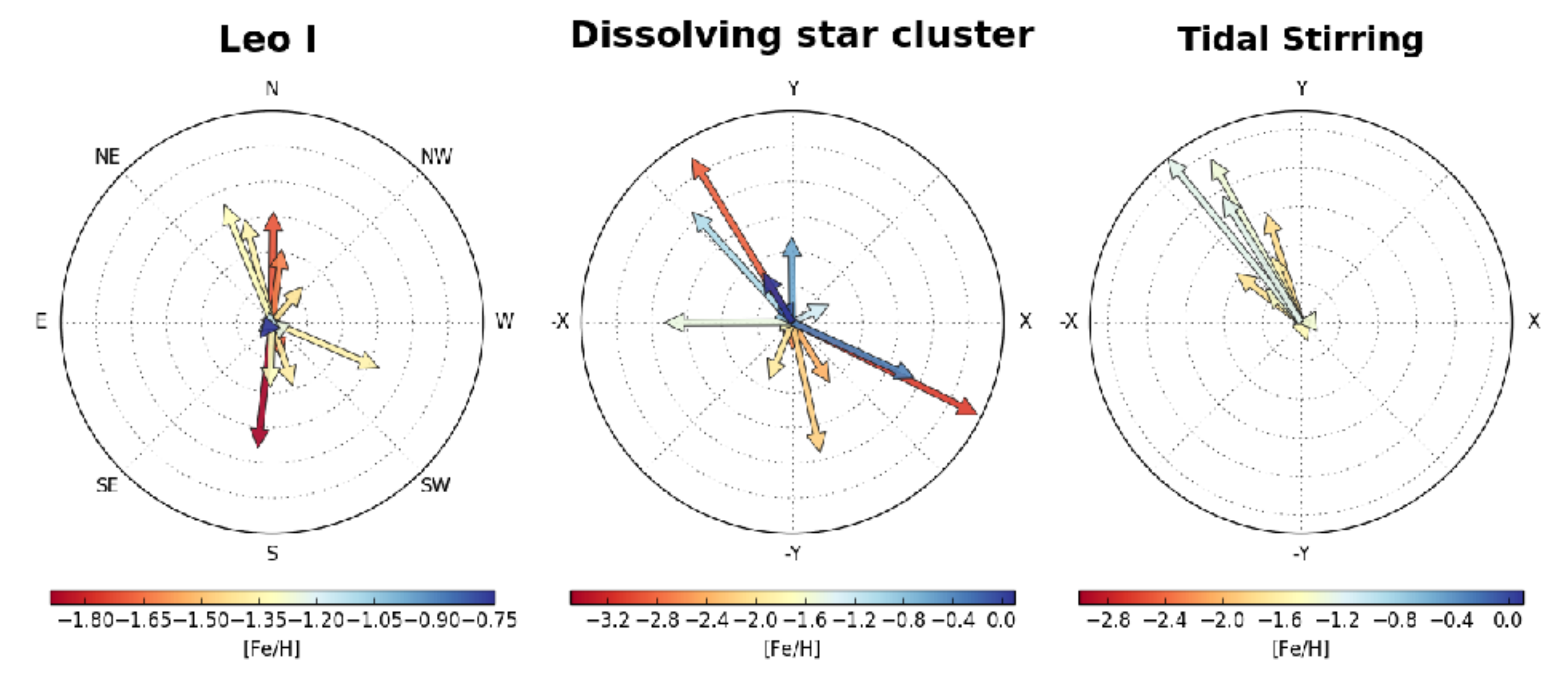}
  \caption{Angular momentum per unit of stellar mass for different
    streams. Left panel: Different angular momentum
    patterns recovered by {\sc Beacon} in the 14 streams of the sample
    of 953 stars from Leo~I,  coloured according to their [Fe/H].
    Middle panel: Angular momentum for a simulation of
    the dissolving star cluster model with 16 star clusters dissolved.
    Circles have radius from $0$~pc$^2$\,s$^{-1}$ to $3 \times
    10^3$~pc$^2$\,s$^{-1}$ with step of $0.5 \times 10
    ^3$~pc$^2$\,s$^{-1}$ for each dotted concentric circle.  Right
    panel: Angular momentum of the streams recovered by {\sc
      Beacon} in a tidal stirring simulation.  There is no net rotation signal in the Leo~I
    observational data and in the dissolving star cluster model, while
    in the tidal stirring model the angular momentum vectors point in
    roughly the same direction, suggesting a remnant rotation signal.} 
  \label{fig:momentums}
\end{figure*}

We can read the results in two fashions.  First, the choice
of the mean velocity is much more important than we
believed before and the number of detected streams drops significantly
by changing the velocity by just $0.01$~km\,s$^{-1}$, while the errors
of every single velocity measurement is of the order of
$2.5$~km\,s$^{-1}$.  Because {\sc Beacon} is based on {\sc Optics}, which is
a purely geometrical clustering algorithm, changing the velocity by just
$0.01$~km\,s$^{-1}$ can change the way clusters are built.  This could
cause some groups to change their member stars and to not fulfil the MCS
and the maxima ratio conditions, and thus to stop being considered by 
{\sc Beacon} as a group.

The second interpretation is that  we see the influence of the randomness of the false detections.
This would make the 14 detections in our study just  outliers;   a
more correct result, looking at the variations of detected streams,
would be $10 \pm 3$ streams.  This would place our result still above
the data points from the pure noise false detections, but with
strongly overlapping error bars, similar to the calculated mean value
from our simulations of the dissolving star cluster models.

Our general result would still stand that the observational data of
Leo~I is in agreement with the DSCM, but with not enough real
streams to have a detection that is significantly above the noise of
having no coherent motion at all.  More data with smaller
observational errors would be necessary to validate or reject this
particular formation scenario for Leo~I.

\subsection{Tidal stirring model}
\label{sec:tsm}

Figure~\ref{fig:momentums} (left panel) shows the projected angular
momentum \textbf{L} of the 14 streams recovered by {\sc Beacon} compared
with the 16 streams of a simulation of the dissolving star cluster
model (middle panel).  According to the dissolving star cluster model
the projected angular momentum is  randomly distributed in the
galaxy, as  shown in Fig.~\ref{fig:momentums}.  The presence of these
streaming motions could be a hint to corroborate our model of the
formation of dwarf spheroidal galaxies.  

Finally, we performed N-body simulations of the tidal stirring and ram
pressure model \citep{Mayer2007}, in which a small disc galaxy is
orbiting a large galaxy like our Milky Way and gets distorted via ram
pressure and tidal stirring.  Ram pressure removes the gas from the
dwarf and tidal stirring destroys the stellar disc and parts of the
DM halo.
We use the code {\sc Gadget2} \citep{Springel2005} to run simulations
of the tidal stirring model using the same set-up as \citet{Mayer2007}.
We vary the initial distance of the disc galaxy to mimic the formation
of a dSph galaxy like Leo~I.  More details of the set-up will be presented in
Alarcon et al.\ (in prep.). 

Preliminary results of the simulations show that after 10~Gyr of
evolution the final objects can have properties that are similar to those 
of the observed dwarf spheroidal galaxies.  For example the spherical shapes, 
exponential surface brightness profiles, and high velocity dispersion values 
which remain constant independent of the distance from the centre, as 
predicted by the model.
Applying {\sc Beacon} with the same parameters used in our analysis of
Leo~I, selecting the same number of stars, and giving the same
metallicity distribution of Leo~I to the particles, we can see that the
final object still has a remnant rotation signal that {\sc Beacon} is
able to recover (see Fig.~\ref{fig:momentums}, right panel).

The strength of this remnant rotation is highly dependent on the
number of orbits around the large galaxy, but if the disc galaxy is
far from its host it will not be able to lose its rotation
completely. We can see in the right panel of Fig.~\ref{fig:momentums}
that the recovered streams are pointing in the same direction.
In this simulation during 10~Gyr the disc galaxy is able to make one
and a half orbits around the large galaxy,  and does not have  enough time to
get rid of its rotation.

A detailed analysis with {\sc Beacon} of simulations using the tidal
stirring model will be performed in the future, and is not part of the
present study.  We show the right panel of Fig.~\ref{fig:momentums}  just to
give a hint of future results.

\section{Discussion and conclusions}
\label{sec:disc-conc}

Using {\sc Beacon} to search for streaming motions arising from the
Dissolving Star Cluster Model in case of perfect data (i.e. no errors
and large sample sizes), almost all streams are detected.  Only 
streaming motions almost perpendicular to the line of sight are very
hard to detect.  A slow rotation pattern close to the centre of
the dwarf galaxy may also escape detection.  In these cases the correct
determination of the dwarf galaxy's radial velocity might become
crucial.  Very eccentric orbits of streams might also not get grouped
together  because the software is searching for rotation
patterns rather than any streaming motion.

False positives decline rapidly and exponentially with the required
minimum cluster size (MCS), so that on perfect simulation data (see
above) they do not play a role. 
This picture changes when adding the typical observational errors (i.e.\
noise) to the data and reducing the sample size dramatically.  In our
study we adopted a sample size of 1000 data points and similar errors
to those associated with the observations of the Leo~1 data.

At small MCS, the real streams get drowned in a sea of false
positives,  but at one point it seems that most false positives get
suppressed and surpassed by real streams.  Otherwise it is not easy to
explain why, in a comparison between no-stream data and
pure-stream data, that we see the data points of the pure-stream data just above
the no-stream data, implying that maybe just 25\% of the streams
present are detected and most of the detected streams are false
positives.  Even so, we see a detection rate of 70\% of the streams
present.  
A possible explanation could be that the algorithm includes a
re-grouping of detections with similar properties (i.e.\ close in
position and/or radial velocities).  Somehow, this favours real streams
over false positives.

Once MCS is increased further and gets closer to about 25\% of the
actual data points belonging to a real stream, the detection rate
drops back into the noise level of false positives.
This means there is a sweet spot or region in MCS where  
the number of detections is expected to be  larger, if real streams are
present, than the number of pure false
positives would predict.  This might not be on a statistically
significant level, but it is visible in Fig.~\ref{fig:streamspleosim}.
Furthermore, in that region of parameter space more detections might
be real than  a simple comparison with no-stream data (i.e.\
pure false positive data) would imply.
 
Putting our results into context, we are not the first  to use {\sc
Beacon} to search for chemo-kinematical patterns in dSph galaxies.
\citet{delPino2017} use {\sc Beacon} to search for chemo-kinematical
patterns in the Fornax dSph galaxy using 2562 stars from different
catalogues \citep{Pont2004, Battaglia2006, Battaglia2008, Walker2009,
  Kirby2010, Letarte2010}.  We made a similar analysis using data from
Leo~I, with a total of 953 stars with their respective metallicity and
velocity values along the line of sight taken from the Keck and Magellan
Telescopes. A direct comparison of the  results  of the two studies are strikingly similar.

\citet{delPino2017} found that approximately $40$\% of the stars may
be part of different substructures, with $985$ stars classified into
$24$ possible circular streams using a ${\rm MCS} = 13$.  This could
indicate that Fornax is a rather complex system with various rotating
components and its spheroidal shape is the superposition of stellar
components with distinct kinematics (see their figure~13).  In our
study we find $14$ possible streams in Leo~I, consisting of about
$40$~\% of all stars in our sample.

If we compare our Fig.~\ref{fig:streamspleosim} with their Fig.~5 we
clearly see the same exponential decrease in false detections as we
have proven with our mock data.  In their analysis we can see the same
signal of possible real streams above the noise level at values
of MCS between 20 and 35.  In their Fig.~6 we see a  peak similar to the one 
in our Fig.~\ref{fig:vhelvsNstream}, even though they use a much
coarser grid of possible velocities as they check for the 2D position
of the galaxy centre as well. 

The star formation history of Fornax together with the spatial
distribution of its stars and their kinematics suggest that Fornax
could be the remnant of a merger between  two 
small galaxies that   collided and merged to form the dSph we
observe today (Amorisco et al 2012, del Pino et al. 2015, del Pino et
al. 2017).  Apart from SFH or chemistry, the kinematic features that are similar
to those observed in Fornax could be explained by the dissolving star
cluster model without the necessity of a merger history (see their
Figs.~9 and 15 and compare them with our Fig.~\ref{fig:momentums}). 

We compare the observational results to mock data deliberately
designed to not show any rotation or streaming signal, to assess the
level of possible false detections in a given data set.  Unluckily,
the size and accuracy of the  present-day data    may not allow us to obtain
streaming motion signals that are statistically significant 
beyond any doubt. 

We find multiple rotating populations in Leo~I with random orientation
of the projected angular momentum, which could indicate that the
overall structure of dSph galaxies is due to the superposition of
different stellar populations with different streaming orbits, as
predicted in the dissolving star cluster model.  The amount of stars
(about $40$~\%) we find to belong to streaming motions is in
agreement with all stars belonging to streams of dissolved star
clusters. 

In the context of this model, these chemo-kinematic patterns are one
of the main predictions as explained and demonstrated in
\citet{Assmann2013b},\citet{Assmann2013a}, and \citet{Alarcon2018}
and come naturally with this model.  However, these patterns could be 
explained by other models assuming rotating progenitors (e.g.\ the
tidal stirring model), which explains the formation of dSph as interactions
between a disc galaxy and a MW size galaxy as well.  These
interactions could be responsible for the reshaping of the disc galaxy
into a dSph \citep{Mayer2010}.  Simulations show that after a
significant time the final object conserves some signatures of this
process and could have remnant rotation around the minor axis.

Another option to explain these patterns is invoking that the
evolution of dSph galaxies could involve several mergers between two
or more smaller galaxies.  In these models two disc galaxies collide,
and as a result they could form a dSph galaxy.  This has been studied before,
and shows that some of the orbits of stars of the progenitors could
remain after the collision \citep{Amorisco2012}.

\citet{Cicuendez2018} made an analysis of the chemical composition
and velocities of stars from the Sextans dSph and have shown that there are
two different populations that have chemo-kinematic patterns  that
could be a sign of a past material accretion.  They reported a shell
structure for young stars and a ring-like structure for older stars. 
\citet{Karlsson2012}, \citet{Kim2019}, and \citet{Aoki2020}
analysed the data of the Sextans dSph as well, and think they have
found the streaming motion of a dissolved star cluster.

These publications show that the analysis of chemo-kinematical
patterns is indeed possible with modern data sets, even though our
study has shown that  the results may not yet be as statistically
convincing as we would like.  Nevertheless, it shows a clear pathway
of how the large data sets that we have today and the much larger data sets 
of the future should be analysed.

We have explained the theoretical pattern a given streaming signal
should have when appearing above a given noise level of false
detections.  This kind of pattern is clearly visible in our study and in the previous work of \citet{delPino2017}.  This pattern cannot
be explained by pure noise and is for us a clear sign that there is a
streaming motion present in Leo~I.  We hope  with future
observations that the signal will become stronger and be significantly above
the noise level.  The fact that it is present in two different
dSph galaxies and two independent studies is more than promising.

Furthermore, thanks to a rigid comparison with mock data having no
streaming motions and simulations built up by only streams, we have
shown that the number of streams we detect is unlikely to be from a
data set without any streaming motion at a 98 to 99\% level of
confidence.  Therefore, even though we cannot tell for certain which
of our detections is based on a real stream and which not, the
evidence points to the presence of streaming motion in the
observational data.
 
In order to have a better understanding of the origin of dSph galaxies, it is
necessary to apply {\sc Beacon} to other dSph galaxies for which high quality
data sets are available in order to search for chemo-kinematical
substructures in those systems to be able to distinguish between the
different formation scenarios for dSph galaxies 
\citep{Diemand2008, Hamuy1996, Koposov2009, Lokas2008, Olszewski1996,
  Perlmutter1999, Riess2004, Spergel1993, Tolstoy2009, Lora2019,
  Valenzuela2007, VanDenBerg2000, Mcconnachie2010}.

We show that Leo~I and Fornax \citep{delPino2017} were very
likely formed by the superposition of different stellar components,
rotating in the galaxy with projected angular momentum pointing in
different directions.  This result is in agreement with the dissolving
star cluster model.  

\section*{Acknowledgments}

AA acknowledges financial support from Carnegie Institution for
Science with its Carnegie-Chile Fellowship, Fondecyt regular
No.1180291 and Centro de Astrofisica y Tecnologias Afines (CATA) grant 
PFB-06/2017.  MF acknowledges funding through Fondecyt regular
No.1180291, Conicyt PII20150171, Quimal No.170001 and the ANID BASAL
projects ACE210002 and FB210003.  AdP acknowledge the financial
support from the Spanish Ministry of Science and Innovation and the
European Union - NextGenerationEU through the Recovery and Resilience
Facility project ICTS-MRR-2021-03-CEFCA.  We thank Marla Geha for
providing the stellar velocities from the Keck/DEIMOS data set.


The data underlying this article will be shared on reasonable request 
to the corresponding author.


\bibliographystyle{aa}
\bibliography{biblio}

\label{lastpage}
\end{document}